\documentclass[amsmath,amssymb,aps,prl,reprint,floatfix]{revtex4-1}

\usepackage[colorinlistoftodos]{todonotes} 
\usepackage{graphicx}
\usepackage{dcolumn}
\usepackage{bm}
\usepackage[none]{hyphenat}
\usepackage[mathlines]{lineno}
\usepackage{float}
\usepackage [english]{babel}
\usepackage [autostyle, english = american]{csquotes}

\newcommand{\ket}[1]{\left|#1\right\rangle}
\newcommand{\bra}[1]{\left\langle#1\right|}

\newcommand{\ketbra}[2]{\left|#1\right\rangle\left\langle#2\right|}
\newcommand{\rom}[1]{\uppercase\expandafter{\romannumeral #1\relax}}

\MakeOuterQuote{"}

\begin{document}

\title{Direct reconstruction of the quantum master equation dynamics of a trapped ion qubit }

\author{Eitan Ben Av}
\email{eitan.ben-av@weizmann.ac.il}\thanks{equal contribution}
\author{Yotam Shapira}\thanks{Equal contribution}
\author{Nitzan Akerman}
\author{Roee Ozeri}
\affiliation{%
Department of Physics of Complex Systems, Weizmann Institute of Science, Rehovot 7610001, Israel
}%

\date{\today}

\begin{abstract}
The physics of Markovian open quantum systems can be described by quantum master equations. These are dynamical equations, that incorporate the Hamiltonian and jump operators, and generate the system's time evolution. Reconstructing the system's Hamiltonian and and its coupling to the environment from measurements is important both for fundamental research as well as for performance-evaluation of quantum machines. In this paper we introduce a method that reconstructs the dynamical equation of open quantum systems, directly from a set of expectation values of selected observables. We benchmark our technique both by a simulation and experimentally, by measuring the dynamics of a trapped $^{88}\text{Sr}^+$ ion under spontaneous photon scattering.
\end{abstract}

\pacs{Valid PACS appear here}
\maketitle

The evolution of open quantum systems, which are coupled to a memory-less bath, are described by the Lindblad master equation, ${\dot\rho\left(t\right)}=\mathcal{L}\left[\rho\left(t\right)\right]$ \cite{GKS1976,Lindblad1976}, where $\rho\left(t\right)$ is the system's density operator and $\mathcal{L}$ is the Lindbladian. The equation generates dynamics due to the system Hamiltonian, and also due to "jump-operators" which encode the coupling between the system and the environment.

At fixed times, the evolution of open quantum systems can be represented by the process matrix, which maps initial density matrices, $\rho\left(0\right)$, to final density matrices, $\rho\left(t\right)$. The process matrix can be experimentally reconstructed by quantum process tomography (QPT) \cite{Chuang1997,Eisert2019}. QPT is often used for computing the process fidelity with respect to some desired quantum process and identify different error channels \cite{Rieve2006,Shabani2011,Rodionov2014,Navon2014}, but it does not characterize the system time-dynamics.

In contrast, reconstruction of the quantum dynamical equations, i.e reconstructing the Hamiltonian and jump operators, allows for decomposition of the different physical mechanisms responsible for the overall evolution. Thus, it allows for prediction of the system state at any time. For Markovian systems, the reconstruction of the full time-dynamics serves as a better tool for analyzing and optimizing systems. 

Here we propose a method for reconstruction of the system's Lindblad master equation, under the evolution of a time-independent Hamiltonian and different decoherence channels, induced by spontaneous scattering of photons by a single $^{88}\text{Sr}^+$ trapped ion-qubit. We focus on three different decoherence channels: amplitude damping, depolariztion, and deploarization accompanied by a coherent rotation. Our measurements provide the first direct reconstruction of the optical Bloch equations \cite{CohenTannoudji1992}.

The dynamics of the trapped ion, after tracing over the scattered photon degrees of freedom and eliminating the short-lived excited states, is reduced to the dynamics of $5S_{\frac{1}{2}}$ ground state Zeeman-qubit coupled to a memory-less environment, we show that, combined with a high-fidelity preparation and measurement, the reconstruction error is dominated by quantum projection noise.
\begin{figure}
    \centering
    \includegraphics[width=\linewidth]{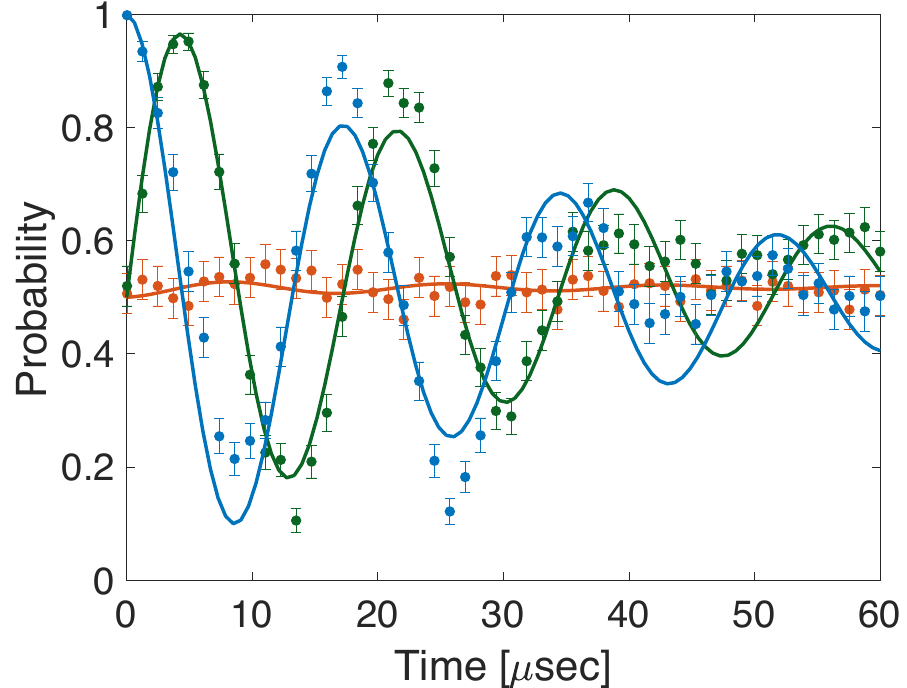}
    \caption{\textbf{Experimental data and corresponding reconstruction of a single trapped-ion spin.} The spin is coherently rotated and coupled to a depolarizing channel (further information below). The data (filled-circles) is obtained by initializing the spin in the $\ket{\uparrow}$ state and measuring it after various evolution times, $t$. The plot shows the probability to measure the $+1$ eigenvalue of the $\sigma_x$ (red), $\sigma_y$ (green) and $\sigma_z$ (blue) Pauli operators. The corresponding values of the resulting reconstruction (solid lines) fit well to the experimental data. Both coherent oscillations and incoherent decay is observed.}\label{fig:projection_vs_reconstruction}
\end{figure}

Dynamical reconstructions have been considered theoretically \cite{Buzek1998,Corry2003,DiFranco2009,Zhang2014,Bairey2018,Bairey2019}, with various assumptions on the allowed dynamics (e.g closed systems, local dynamics etc.). Furthermore, there have been experimental demonstrations of a Hamiltonian reconstruction \cite{deClercq2016} and of an open-system dynamical reconstruction \cite{Howard2006}. In the latter, the reconstruction was performed by piecing together a sequence of independent QPTs.

Here we directly optimize an estimated Lindbladian by using a cost function that compares our measured data with corresponding data that is numerically generated by our estimation. The resulting reconstructed Lindbladian is then optimal for all measured quantities at all measurement times. Figure \ref{fig:projection_vs_reconstruction} shows an example of such a reconstruction, due to a series of measurements on a single trapped-ion qubit. The spin is coherently rotated and coupled to a depolarizing channel (further information below). The measured data (filled-circles) of different observables is compared to their expectation values (solid lines) predicted by the reconstructed Linbladian, showing a good fit.

The evolution of any quantum system coupled to a memory-less environment, is described by the Lindblad dynamical equation, 
\begin{equation}
\begin{split}
\dot{\rho}\left(t\right)=&-i[H,\rho\left(t\right)]+\sum_{n=1}^{N^{2}-1}\gamma_{n}\mathcal{L}_{n}[\rho\left(t\right)]\equiv\mathcal{L}\left[\rho\left(t\right)\right]\\
\mathcal{L}_{n}[\rho\left(t\right)]=&L_{n}\rho\left(t\right)L_{n}^{\dagger}-\frac{1}{2}\left\{L_{n}^{\dagger}L_{n},\rho\left(t\right)\right\},
\end{split}\label{eqnLindblad}
\end{equation}
where $H$ is the system Hamiltonian, the $\gamma_n$'s are decoherence rates, $L_{n}$ are the jump operators, and $N=2^n$ is the dimension of a $n$-qubit Hilbert space. Here, and in what follows, we use $\hbar=1$. We note that, similar to Schr\"odinger's equation, this is a linear equation, which can thus be denoted by a single linear (super-)operator, the Lindbladian $\mathcal{L}$. This trace-preserving, completely positive, operation is then described by $16^n-4^n$ degrees of freedom, its exact form can be found in ref \cite{Corry2003}.

Equation \eqref{eqnLindblad} is formally solved by exponentiation of the Lindbladian,
\begin{equation}
\rho\left(t\right)=\mathcal{T}\left[e^{\int_0^t\mathcal{L}\left(t^\prime\right)dt^\prime}\right]\rho\left(t=0\right),\label{eqnRhoOft}
\end{equation}

where $\mathcal{T}$ is the time-ordering operator. In this study we treat time-independent systems, for which the time-ordering can be dropped and the integration is trivial. The operator $\mathcal{L}$ contains all the information about the dynamics of the system. Therefore, reconstructing the dynamics is equivalent to obtaining $\mathcal{L}$.

In principle, $\mathcal{L}$ can be reconstructed by taking the logarithm of $\rho\left(t\right)$ at a fixed evolution time. However, this "inverse" reconstruction is unstable in the sense that small measurement errors in $\rho$, which are inherent in any tomographic method, can result in unbounded errors in the estimation of $\mathcal{L}$, making the problem ill-conditioned \cite{Leroy1987,Tarantola1987}. 

Therefore, an alternative approach is required. Specifically we look for a method where an estimated Lindbladian is guessed out of the space of valid processes and iteratively optimized. This is performed as follows, the experimental system is prepared in well-defined initial states, and measured after multiple evolution times. These measurements are compared to a calculation of the corresponding expectation values, after evolution of the initial state given by Eq. \eqref{eqnRhoOft}. A suitable Lindbladian is chosen by minimizing the difference between the two results, evaluated for various initial states, evolution times and observables. 

Specifically, we initialize the system to one of $K$ fiducial states, $\left\{\ket{\psi_k}\right\}_{k=1}^{K}$, evolve it to time $t$ (out of a sequence of times in $[0,T]$) and evaluate one of $B$ observables, $\left\{O_b\right\}_{b=1}^{B}$. The measurement results are distributed according to,
\begin{equation}
    P_{b,k,t}^\mathcal{L}\left(j\right)=\bra{j_b}e^{\mathcal{L}t}\left[\ketbra{\psi_k}{\psi_k}\right]\ket{j_b},\label{eqnDist}
\end{equation}
with $\ket{j_b}$ corresponding to the $j$'th eigenvector of $O_b$ such that $\sum_{j}P_{b,k,t}^\mathcal{L}(j)=1$.

By performing $M$ identical measurements and computing the relative recurrence of the different outcomes we obtain a probability distribution $P_{b,k,t}(j)$. In the presence of quantum projection noise $\left|P_{b,k,t}(j)-P_{b,k,t}^\mathcal{L}\left(j\right)\right|\propto1/\sqrt{M}$.

The optimization then minimizes the sum of "distances" between the distribution $P_{b,k,t}\left(j\right)$, and its reconstructed estimate, $\hat{P}_{b,k,t}\left(j\right)$ via the cost function, 
\begin{equation}
    C=\sqrt{\frac{1}{N}\sum_{b,k,t}\left[d\left(\hat{P}_{b,k,t},P_{b,k,t}\right)\right]^2}+\varepsilon\left(\hat{\cal{L}}\right), \label{eqnCost}
\end{equation}

Where $N$ is a normalization such that $\sum_{b,k,t}1=N$,  $\hat{P}_{b,k,t}\left(j\right)$ is inferred from the current estimation of $\mathcal{L}$ using Eq. \eqref{eqnDist}, and $d\left(x,y\right)$ is a pre-metric \footnote{A pre-metric is a generalization of a distance function. It satisfies, $d(x,y)\ge0$ and $d(x,y)=0$ i.f.f $x=y$. That is, it is in general not symmetric, and does not satisfy the triangle inequality}, defined on the probabilities $x(j)$ and $y(j)$.

In practice, we choose the Kullback-Leiblar divergence as our pre-metric \cite{Kullback1951}, as it has shown numerically to yield favorable results. That is, we set $d_{\text{KL}}\left(x,y\right)=\sum_{j=1}^{j=J} x\left(j\right)\log\frac{x\left(j\right)}{y\left(j\right)}$, where $J$ is the number of possible outcomes. For spin-$\frac{1}{2}$ $J=2$.

We have included a penalty function to Eq. \eqref{eqnCost}, $\varepsilon\left(\hat{\cal{L}}\right)$, which is used to impose a-priori constraints on the reconstruction. Here we use it in order to penalize reconstructions with rates that are faster than the sampling rate, i.e excluding processes which oscillate faster than the Nyquist frequency. 

For a single qubit it is convenient to choose the Pauli matrices as the measurement basis $O_b=\sigma_b$ with $b=x,y,z$, and the fiducial states $\left\{\ket{0},\ket{1},\ket{+}=\frac{\ket{0}+\ket{1}}{\sqrt{2}},\ket{i}=\frac{\ket{0}+i\ket{1}}{\sqrt{2}}\right\}$. For a single qubit there are at most two measurement results, so we can drop the $j$ index. The resulting 12 time-series $P_{b,k,t}$ are linearly-independent and sufficient to reconstruct the Lindbladian. 

A single qubit process has an appealing geometric interpretation on the well-known Bloch sphere, for which all pure states reside on the sphere surface and all mixed states reside within its volume. The system evolution is then a "movie" in which the Bloch sphere, which represents all initial system states, continuously rotates and deforms. 

Hamiltonian rotations map pure states onto pure states, and require three degrees of freedom; two for choosing the rotation axis and one for the rotation angle. The jump operators can be decomposed to dilation and displacement of the Bloch sphere. For dilation, three degrees of freedom specify an orthogonal Cartesian system and three other degrees of freedom specify stretches along each orthogonal direction. Displacement by a vector makes use of additional three degrees of freedom. In total, a single qubit Lindblad equation is determined by 12 parameters.

To benchmark our method we randomly choose 10,000 processes, and simulate measurement results in the presence of quantum projection noise. We then employ our reconstruction method. The reconstruction error is evaluated as, 
\begin{equation}
    \varepsilon_\text{r}=\left\Vert \cal{L}_{\text{original}}-\cal{L}_{\text{estimate}}\right\Vert_\text{F},
    \label{eqnReconError}
\end{equation}

where ${\cal{L}}_{\text{original}}$ is the randomly chosen Lindbladian, and ${\cal{L}}_{\text{estimate}}$ is its corresponding reconstructed Lindbladian, both in their matrix form (see \cite{Corry2003}), and $\left\Vert X \right\Vert_\text{F}=\sqrt{\text{Tr}(X^{\dagger}X)}$ is the Frobenius operator norm \footnote{A norm between matrices that is invariant under change of basis. For diagonalizable matrices it is equal to $\sqrt{\sum_i\lambda_i^2}$, where the $\lambda_i$'s are the eigenvalues}. 

\begin{figure}
    \centering
    \includegraphics[width=\linewidth]{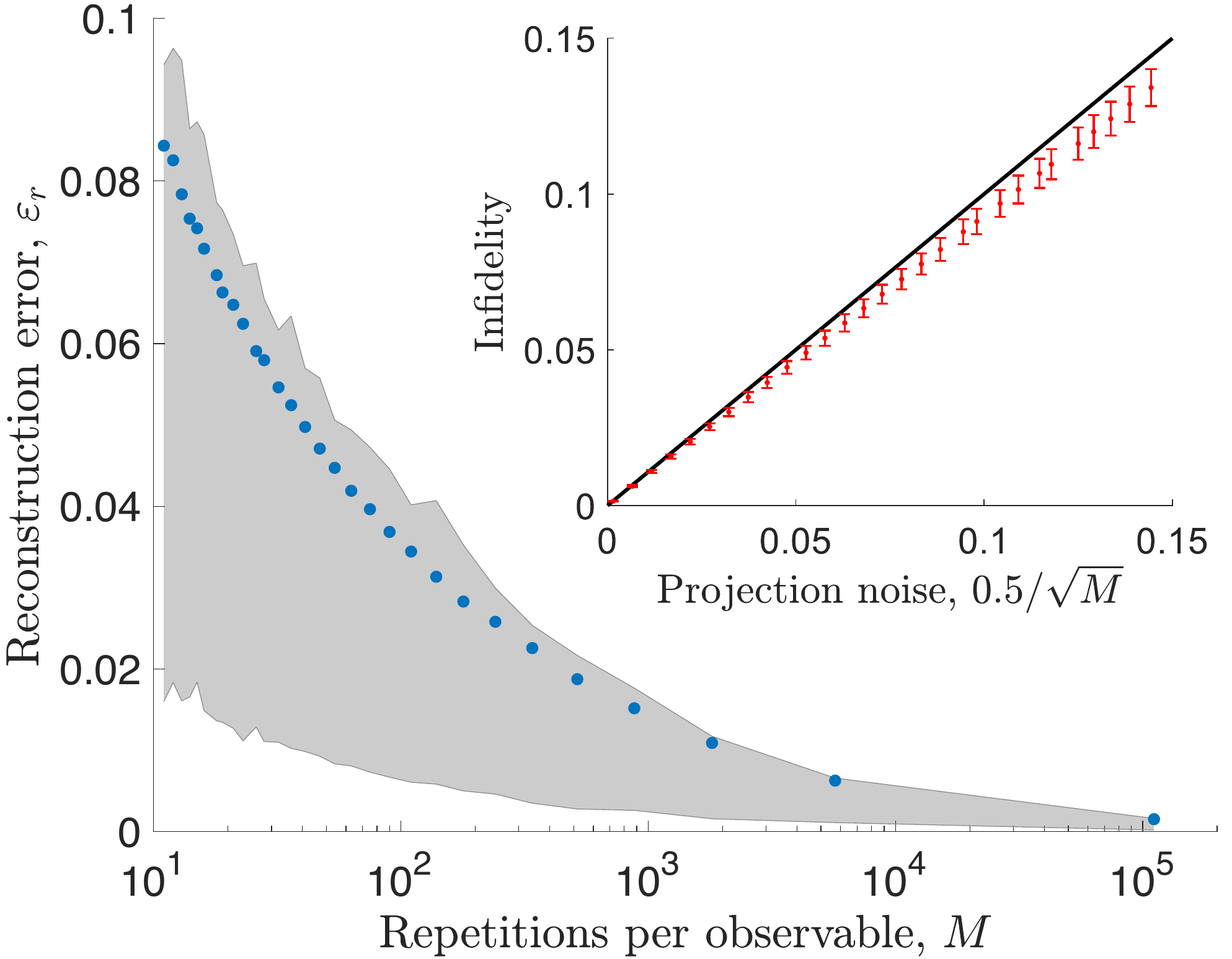}
    \caption{\textbf{A benchmark of our reconstruction method through simulation.} The Reconstruction error, $\varepsilon_\text{r}$, as defined in Eq. \eqref{eqnReconError}, for a varying number of repetitions per observable, $M$ (log scale). Each point corresponds to the average error due to 10,000 randomly chosen independent simulated processes (blue). The shaded gray background marks where $68\%$ of the errors of the different processes are. The inset shows the infidelity,  defined in Eq. \eqref{eqnCost} and  \eqref{eqnC2Cost}. Error bars represent one standard deviation. Clearly the infidelity shows a similar behaviour to $\varepsilon_\text{r}$, validating its practical use as a stopping criterion.}
    \label{fig:Rec_dist_vs_pn}
\end{figure}

Figure \ref{fig:Rec_dist_vs_pn} shows the resulting reconstruction error of this benchmark as a function repetitions per observable. As expected the reconstruction error in Fig. \ref{fig:Rec_dist_vs_pn} improves with increasing number of repetitions and reduction of quantum projection noise.  

Since experimentally the original Lindbladian is unknown, the reconstruction error in Eq. \eqref{eqnReconError} is not accessible, and therefore cannot serve as a stopping criteria for the optimization iterations. Therefore, we also use the numerical benchmark to evaluate the infidelity of reconstruction after the numerical search has concluded. We define the reconstruction infidelity using the expression in Eq. \eqref{eqnCost}, this time with  the $C_2$ metric, i.e,
\begin{equation}
    d_{C_2}\left(x,y\right)=\sqrt{\frac{1}{J}\sum_{j=1}^J\left[x\left(j\right)-y\left(j\right)\right]^2},\label{eqnC2Cost}
\end{equation}
defined on the probabilities $x(j)$ and $y(j)$. This metric is helpful since with it, the infidelity is simply the root mean square (RMS) of the differences between the reconstructed distribution and the measured $P_{b,k,t}$ (e.g the vertical distance between the points and solid lines in Fig. \ref{fig:projection_vs_reconstruction}). As such, for an ideal reconstruction the resulting infidelity is the RMS of the measurement noise. In the case of a spin-$\frac{1}{2}$ with quantum projection noise, the average infidelity is bounded by $\frac{0.5}{\sqrt{M}}$ where $M$ is the number of repetitions per observable.

The inset in Fig. \ref{fig:Rec_dist_vs_pn} shows the infidelity evaluated for the reconstructions above. As seen, the average infidelity is bounded by $\frac{.5}{\sqrt{M}}$ (solid black lines), the maximal projection noise due to $M$ measurements. As the projection noise decreases, both the reconstruction error and infidelity decrease as well, indicating the infidelity is a valid stopping criteria for the reconstruction iterations. 

\begin{figure}
    \centering
    \includegraphics[width=\linewidth]{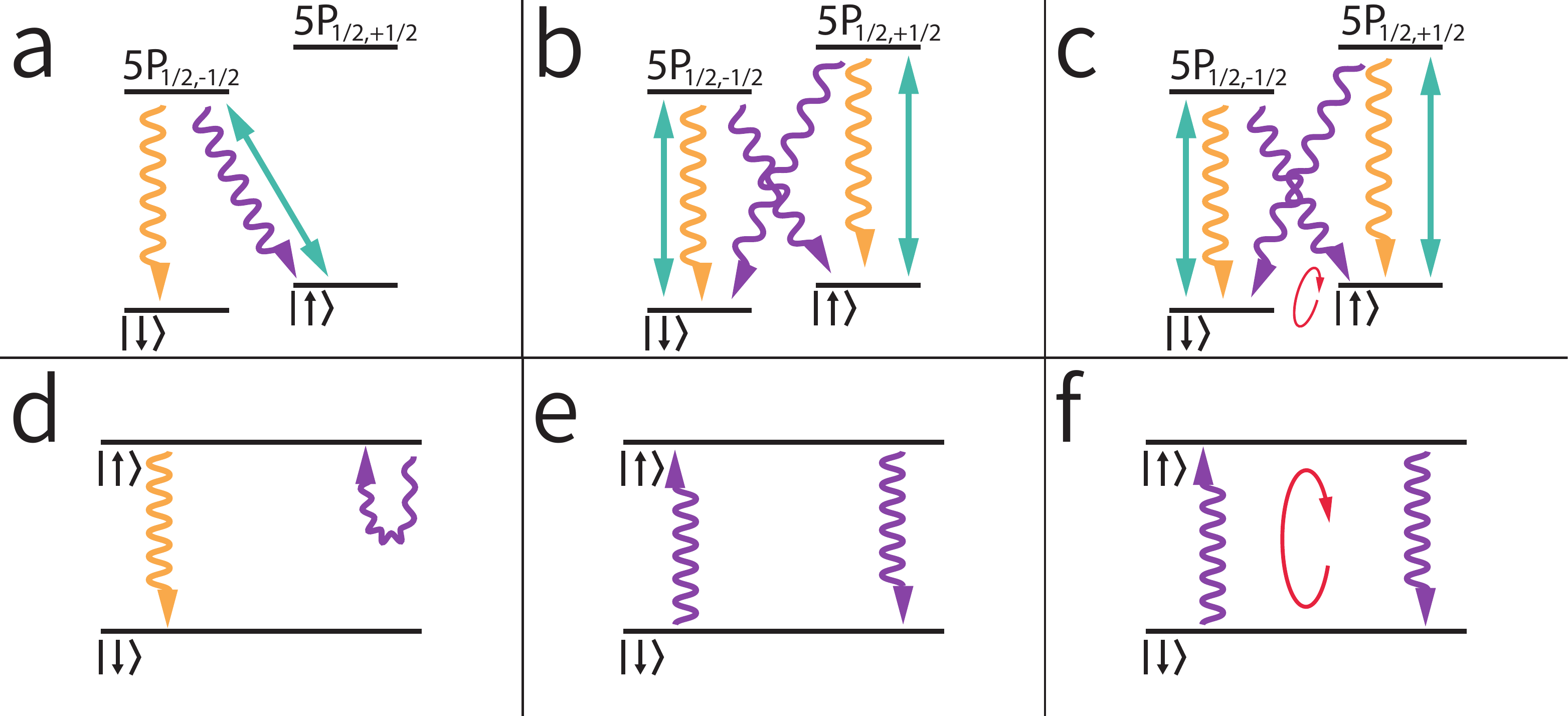}
    \caption{ \textbf{Physical picture and corresponding two-level dynamics.} (a)-(c): Relevant level structure of the $^{88}\text{Sr}^+$ ion, coupling fields and spontaneous decay channels, showing the $5S_{\frac{1}{2}}$ qubit manifold (two bottom levels) and the short-lived $5P_{\frac{1}{2}}$ manifold (two upper levels). The $D_{\frac{3}{2}}$ and $D_{\frac{5}{2}}$ are not shown here since their contribution is negligible. (d)-(f): The corresponding open dynamics induced on the qubit manifold. 
    (a) \& (d) Amplitude damping. A $\sigma^-$-polarized $422\text{ nm}$ laser field (green arrow) optically pumps the $\ket{\uparrow}$ state to the $\ket{\downarrow}$ state, generating a spontaneous decay effect (yellow line) and a decoherence effect (purple line).
    (b) \& (e) Depolarization channel. A $\pi$-polarized $422\text{ nm}$ laser field (green arrows) excites both qubit states to the $5P_{\frac{1}{2}}$ manifold, which decay in a Raman process (purple arrows), generating a decay to the fully mixed state, or in a Rayleigh process (yellow arrows), which leaves the state unchanged.
    (c) \& (f) Depolarization channel, with coherent rotation. The same $422\text{ nm}$ laser field as in (b) \& (e) is used, supplemented with RF coupling between the two qubit levels (red arrow), generating coherent oscillations between the two qubit state.}
    \label{fig:8LS2LS}
\end{figure}

We demonstrated the reconstruction method on a single $^{88}\text{Sr}^+$ ion, trapped in a linear Paul trap. We used the $5S_\frac{1}{2}$ Zeeman manifold as the effective qubit states. Coherent transitions between the two qubit levels are induced by an RF field tuned to the Zeeman transition. The fast decaying $5P_\frac{1}{2}$ and long-lived $4D_\frac{5}{2}$ manifolds are coupled to the qubit states by a $422$ nm and $674$ nm laser fields respectively. These lasers allow for state preparation by optical-pumping and state measurement by state-dependent fluorescence. Due to a $1:14$ branching ratio probability of decay from the $5P_\frac{1}{2}$ level to the $4D_\frac{3}{2}$ manifold, it is re-pumped by a $1092$ nm laser (for further information see \cite{Akerman2012,AkermanPhd}).

The open memory-less dynamics is tailored by using the $S\rightarrow P$ transition, which effectively acts as a Lindbladian in the qubit subspace. For example, an amplitude damping channel (Fig. \ref{fig:8LS2LS}a) is implemented by illuminating the ion with a $\sigma^-$-polarized $422 \text{ nm}$ laser (green arrow). This induces transitions from the $\ket{\uparrow}$ qubit state to the $5P_{\frac{1}{2},\frac{1}{2}}$ state, which then quickly decays back to the qubit subspace, i.e it optically pumps the qubit to the $\ket{\downarrow}$ state. A depolarization channel (Fig. \ref{fig:8LS2LS}b) is implemented by illuminating the ion with a $\pi$-polarized $422 \text{ nm}$ laser, which cycles both qubit states through the $5P_\frac{1}{2}$ manifold. As a result, the $\ket{\uparrow}$ state decays to the $\ket{\downarrow}$ state and vice versa. Using, the above open quantum channels and coherent qubit rotations we implemented three different dynamics: amplitude damping, depolarization, and depolarization accompanied by coherent rotation. Figure \ref{fig:8LS2LS} illustrates the corresponding levels and couplings.

We reconstructed the ion-qubit dynamics using the methods above. We evaluated each $P_{b,k,t}$ by averaging $M=625$ (amplitude damping channel), and $M=200$ (depolarization channels) measurements per observable. This bounds our projection noise per data point by $0.5/\sqrt{M}$, i.e 0.02 for the former and 0.0354 for the latter. The resulting infidelities are $0.008$, $0.016$  and $0.029$ for amplitude damping, depolarization, and depolarization with coherent rotation of the qubit, respectively. 

\begin{figure}
    \centering
    \includegraphics[width=\columnwidth]{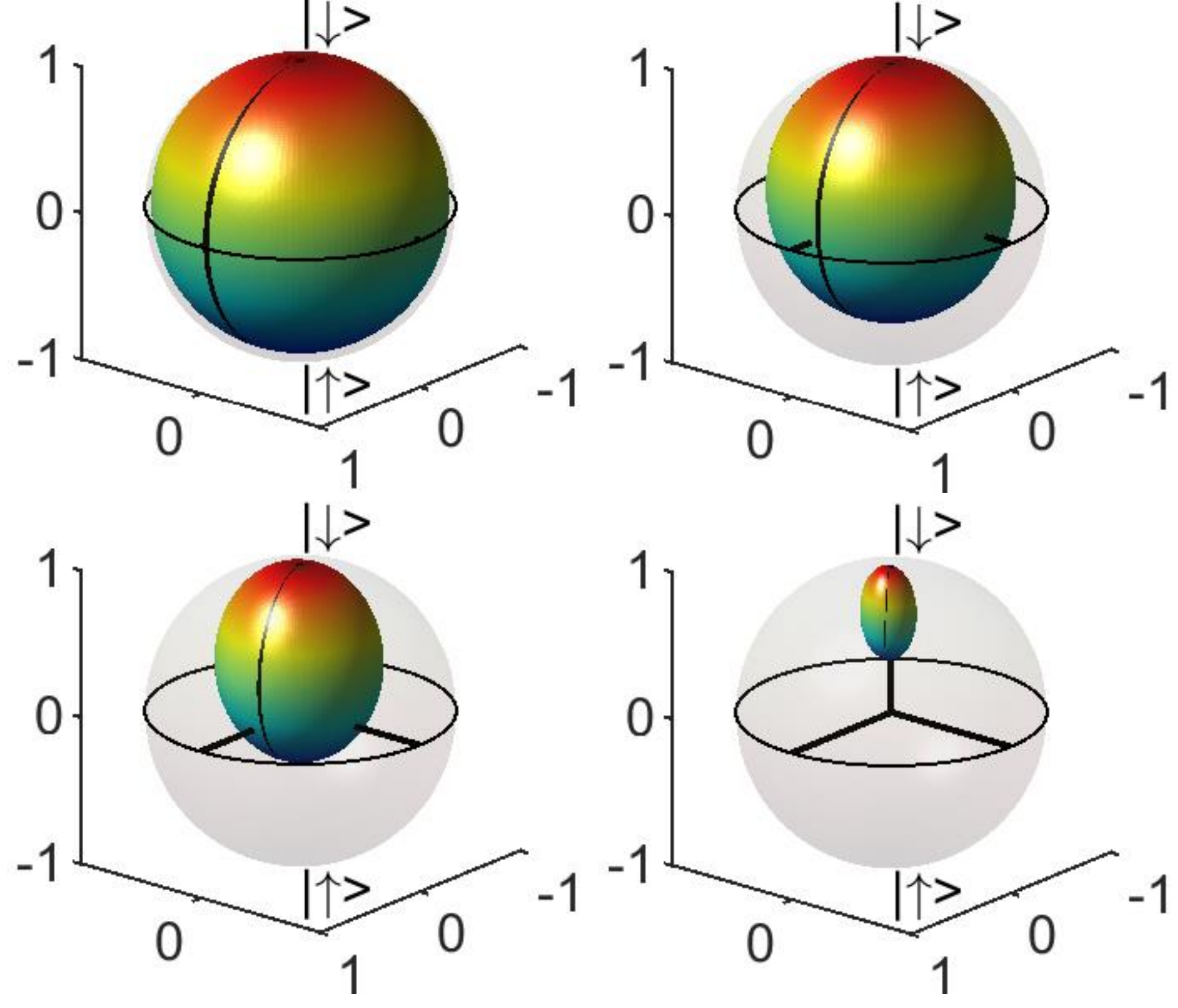}
    \caption{Snapshots of Bloch sphere reconstruction movie for amplitude damping. The Bloch sphere is shown at four different times, $t=0.2, 1, 3, 9 \left[{\mu}\text{s}\right]$. The snapshots order is top-left, top-right, bottom-left and bottom-right. The sphere shrinks to the north pole, indicating that all initial states relax to the $\ket{\downarrow}$ state.}\label{fig:bloch_sphere_movie_1st_senario}
\end{figure}

Using the reconstruction of these three channels, we were able to reconstruct the master equation behind the dynamics we implemented. Using the reconstructed equation we can graphically present this dynamics as a movie of the Bloch sphere of states evolving in time. Figure \ref{fig:bloch_sphere_movie_1st_senario} shows snapshots of the Bloch sphere evolution movie for the case of amplitude damping. On the Bloch sphere this is seen as a deflation towards the $\ket{\downarrow}$ state, represented by the north pole, as expected. The Bloch sphere does not rotate, indicating that there is no unitary Hamiltonian dynamics involved. 

The main jump operators we recover are $\ketbra{\downarrow}{\uparrow}$ and $\ketbra{\uparrow}{\uparrow}$ where the latter has a rate two times larger than the former due to the corresponding Clebsch-Gordan coefficients. A direct solution of the Lindblad equations shows that this results in a dechorence rate which is 1.5 times bigger than the population decay rate. Our reconstruction recovers the ratio $1.55\pm0.15$. On the Bloch sphere this is seen as an elongation in the $\hat{z}$ direction.

\begin{figure}
    \centering
    \includegraphics[width=\columnwidth]{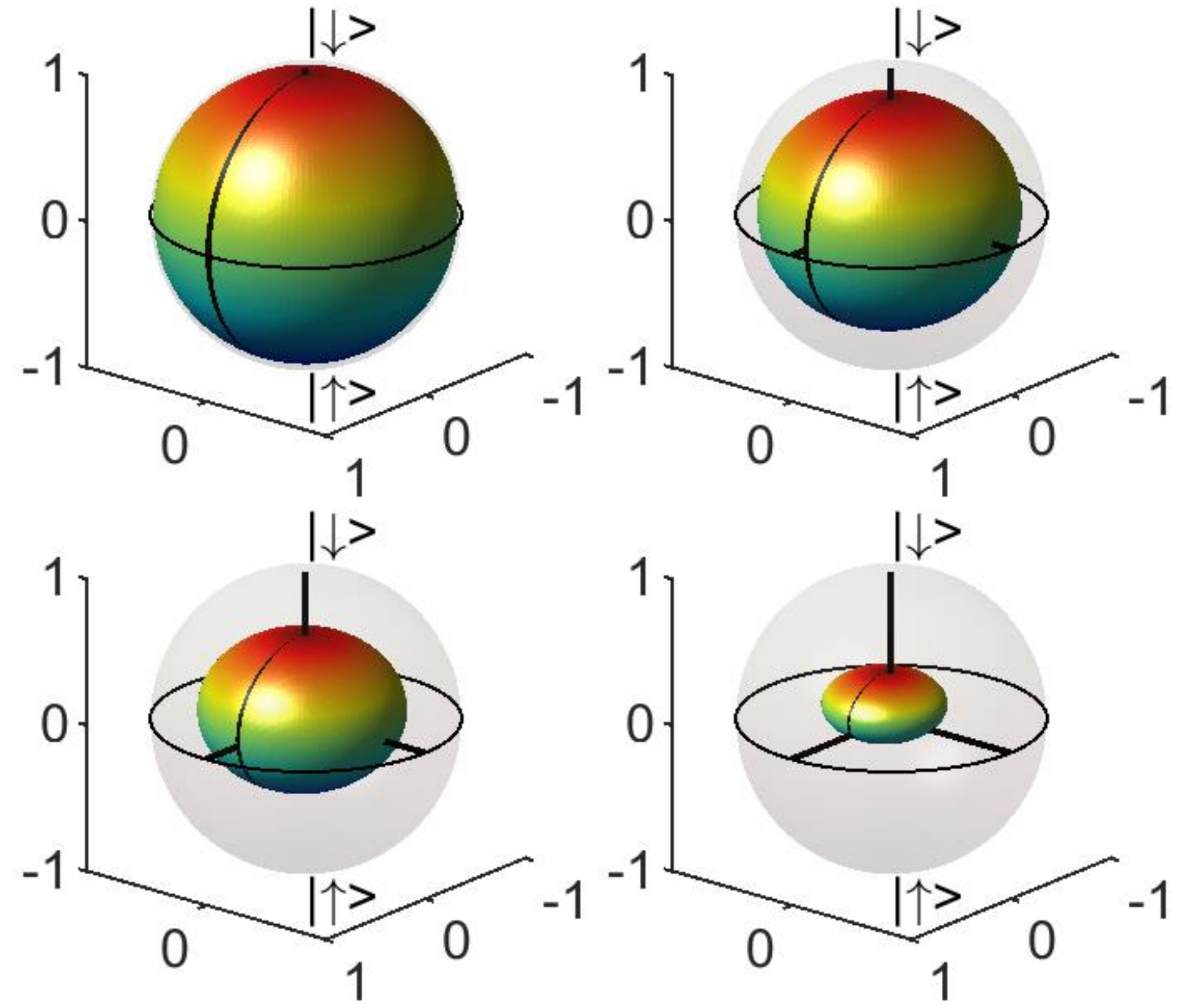}
    \caption{Snapshots of Bloch sphere reconstruction movie for depolarization. The Bloch sphere is shown at four different times, $t=1, 6, 16, 37 \left[{\mu}\text{s}\right]$. Here the sphere shrinks to the origin, indicating all initial states relax to the fully mixed state.}\label{fig:bloch_sphere_movie_2nd_senario}
\end{figure}

Figure \ref{fig:bloch_sphere_movie_2nd_senario} shows the results of the reconstructed master equation in the depolariztion channel. Here we mainly reconstruct the jump operators $\ketbra{\uparrow}{\downarrow}$, and its conjugate (Fig. \ref{fig:8LS2LS}e purple arrows), caused by Raman photon scattering. On the Bloch sphere this is seen as a deflation of the sphere towards the center, corresponding to the limit of a thermal state of an infinite-temperature system.

We note that jump operators of the form $\ketbra{\uparrow}{\uparrow}$ and $\ketbra{\downarrow}{\downarrow}$, i.e Rayleigh scattering operators, do not appear in the dynamics. This is because Rayleigh scattered photons (Fig. \ref{fig:8LS2LS}b yellow arrows) do not contain information about the qubit state in our case. Thus the deflation rate of the sphere in the $\hat{z}$ axis is faster than in the $\hat{x}$ and $\hat{y}$ directions, giving rise to anisotropy in the depolarization process. Such Rayleigh scattering however does contribute a coherent $\sigma_z$ rotation due to a Stark-shift effect \cite{Akerman2012raman,Glickman2013}.

\begin{figure}
    \centering
    \includegraphics[width=\columnwidth]{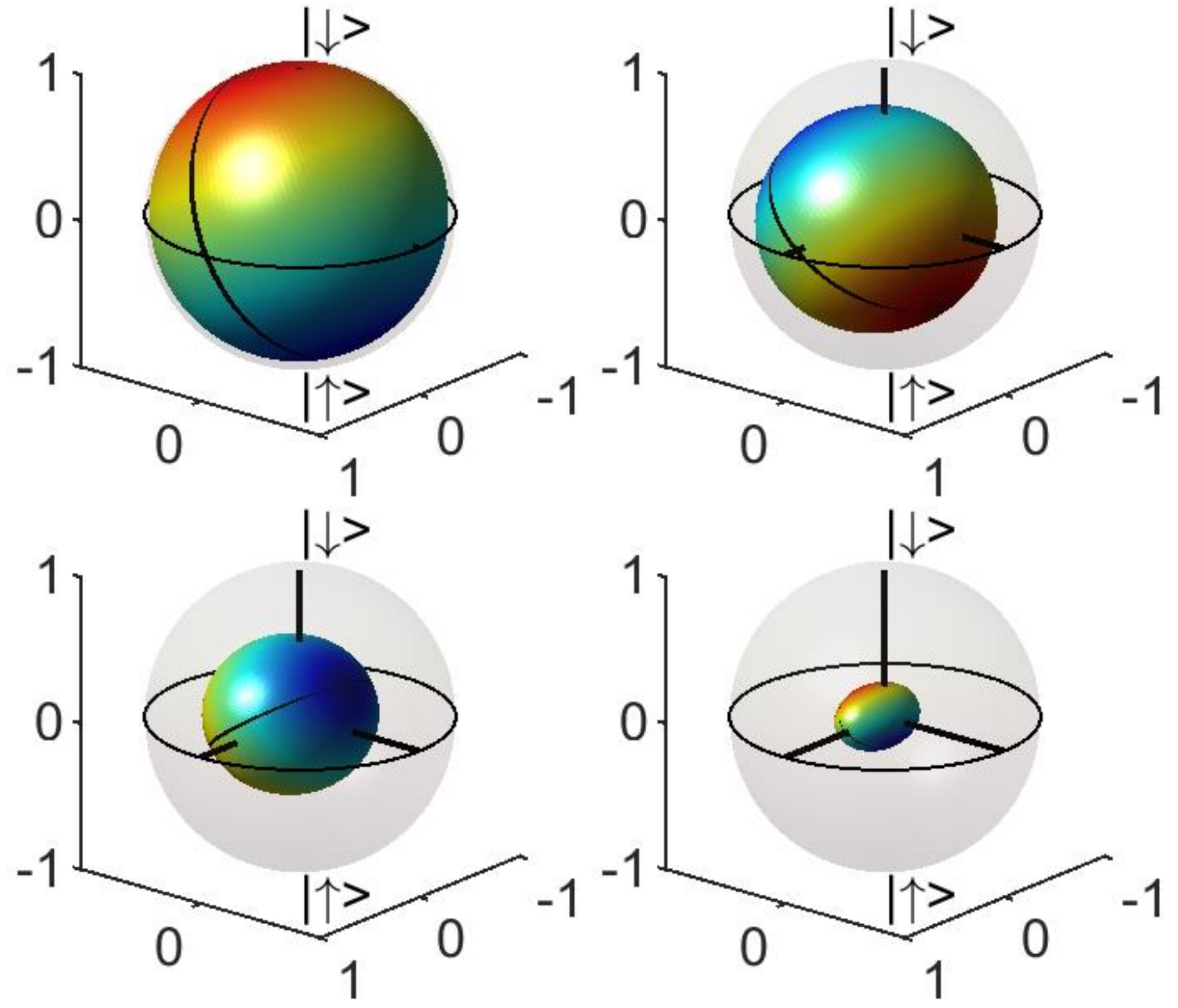}
    \caption{Snapshots of Bloch sphere reconstruction movie for depolarization while coherently rotating the qubit. The Bloch sphere is shown at four different times, $t=1, 10, 23, 53 \left[{\mu}\text{s}\right]$, along the system evolution. Here the sphere shrinks to the origin while rotating around the $\hat{x}$ axis.}\label{fig:bloch_sphere_movie_3rd_senario}
\end{figure}

So far we have only discussed purely non-unitary processes. However, often decoherence occurs during Hamiltonian dynamics. As a simple demonstration, we generalize the anisotropic depolarization channel above by adding a $\sigma_x$ drive. This is implemented by turning on an on-resonance RF field which coherently couples the two qubit states, and, in the absence of decoherence, generates Rabi oscillations between them.

Figure \ref{fig:bloch_sphere_movie_3rd_senario} shows an anisotropic depolarization channel while coherently rotating the qubit. This in general generates both coherent oscillations and decay of the spin.

The resulting evolution on the Bloch sphere is similar to the anisotropic depolarization scenario above, but with an added rotation of the sphere around the $\hat{x}$ axis. This rotation interchanges the $\ket{\uparrow}$ and $\ket{\downarrow}$ with the $\ket{+i}$ and $\ket{-i}$ eigenstates of $\sigma_y$ at the edges of the $\hat{y}$ axis, leading to a fast decay of all four states. The $\sigma_x$ eigenstates decay much slower, as they are eigestates of the rotation. Therefore the corresponding sphereoid is elongated along the $\hat{x}$ axis and gradually becomes symmetric around it. Notably, the previous symmetry around the $\hat{z}$ axis is now broken.

In conclusion, we have presented a general method of reconstructing the Lindblad dynamical equation from sets of observables over time. We used simulations in order to devise a stopping criterion for the reconstruction method, and verified that the reconstruction error is small and comparable to the measurement shot noise. Furthermore, we have demonstrated our method on a trapped $^{88}\text{Sr}^+$ ion in three different open quantum system dynamics channels, implemented using spontaneous photon scattering. Our measurements constitute the first direct reconstruction of the optical Bloch equations \cite{CohenTannoudji1992}. Our method is applicable both for verification of engineered dynamics and for investigation of unknown processes and noise.

\begin{acknowledgments}
This work was supported by the Crown Photonics Center, the Israeli Science Foundation, the Israeli Ministry of Science Technology and Space and the Minerva Stiftung
\end{acknowledgments}

\end{document}